\documentclass{iau}
\usepackage{graphicx} 

\title[IAUS291.~~ARTEMIS with LOFAR international stations] 
{Observations of transients and pulsars with LOFAR international stations and the ARTEMIS backend} 

\author[M. Serylak et al. ]  
{Maciej~Serylak$^{1,2}$,
Aris~Karastergiou$^3$,
Chris~Williams$^4$,
Wesley~Armour$^5$,
Michael Giles$^5$\\
\and the LOFAR Pulsar Working Group}

\affiliation{$^1$Station de Radioastronomie de Nan\c{c}ay, Observatoire de Paris, CNRS/INSU, \\18330 Nan\c{c}ay, France \\email: {\tt maciej.serylak@obs-nancay.fr} \\[\affilskip]
$^2$Laboratoire de Physique et Chimie de l'Environnement et de l'Espace, \\LPC2E UMR 7328 CNRS, 45071 Orl\'eans Cedex 02, France \\[\affilskip]
$^3$Astrophysics, University of Oxford, Keble Road, OX1 3RH, United Kingdom \\ email: {\tt aris@astro.ox.ac.uk} \\[\affilskip]
$^4$Oxford e-Research Centre, University of Oxford, Keble Road, OX1 3QG, United Kingdom \\email: {\tt christopher.williams@oerc.ox.ac.uk}  \\[\affilskip]
$^5$Institute for the Future of Computing, University of Oxford, Keble Road, OX1 3QG, \\United Kingdom \\email: {\tt wes.armour@oerc.ox.ac.uk} \\[\affilskip]
}

\pubyear{2012}
\volume{291}  
\jname{\mbox{Neutron Stars and Pulsars: Challenges and Opportunities after 80 years}}
\editors{J. van Leeuwen, ed.} 
\begin{document}

\maketitle

\begin{abstract}
The LOw Frequency ARray -- LOFAR -- is a new radio interferometer
 designed with emphasis on flexible digital hardware instead of
mechanical solutions. The array elements, so-called stations, are
located in the Netherlands and in neighbouring countries. The design
of LOFAR allows independent use of its international stations, which,
coupled with a dedicated backend, makes them very powerful telescopes in
their own right. This backend is called the Advanced Radio Transient
Event Monitor and Identification System (ARTEMIS). It is a combined
software/hardware solution for both targeted observations and
real-time searches for millisecond radio transients which uses
Graphical Processing Unit (GPU) technology to remove interstellar
dispersion and detect millisecond radio bursts from astronomical
sources in real-time.
\end{abstract}


\firstsection 
\section{Introduction}

The development of general-purpose GPUs, able to perform tasks done
previously by Central Processing Units (CPUs) offers an attractive
alternative for radio-astronomical applications. Because the storage
and off-line processing of such a vast amount of data is difficult and
costly, the new generation of radio telescopes, such as LOFAR,
requires High Performance Computing (HPC) solutions to process the
enormous volumes of data that are typically produced during a survey
for fast radio transients (\cite[Jones
  et~al. 2012]{jwt+12}). Real-time searches for radio transients,
which use GPU technology to remove interstellar dispersion and detect
radio bursts from astronomical sources in real-time are now
possible. Also it is important to note that real-time processing
offers the chance to react as fast as possible and to conduct
follow-up observations of any event. We report here on the
installation of a new backend which can be used with a single
international LOFAR station. The details of the LOFAR and pulsar observations with LOFAR can be found in \cite{sha+11} and in Kondratiev et~al. (these proceedings).

\begin{figure}[t]
\begin{center}
\vspace{3mm}\includegraphics[width=0.95\textwidth]{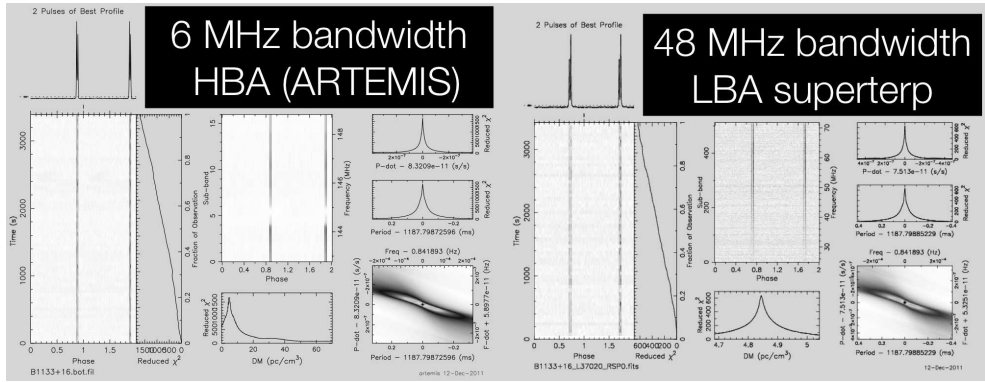}
\caption{Two diagnostic plots made from simultaneous observations done with single international LOFAR station using HBA antennas covering 6 MHz of bandwidth and recorded with the ARTEMIS backend (left) and the LOFAR core at the full observable bandwidth of 48 MHz using LBA antennas (right). This example illustrates that international LOFAR stations could be used together with the LOFAR core for simultaneous observations.}
\label{fig1}
\end{center}
\end{figure}

\section{ARTEMIS -- backend for international LOFAR stations}

ARTEMIS is a combined hardware/software backend, which attaches in a non-disruptive way to the LOFAR station hardware. The hardware consists of 4 12-core servers hosting high-end NVIDIA GPU cards. These servers are all fed data through a broadband (10 Gigabit Ethernet) switch, which is also responsible for sending the data back to the Netherlands during the normal International LOFAR Telescope (ILT) operations. The data being processed adds up to a stream of 3.2 Gbits/s, which consists of a sky bandwidth of approximately 48 MHz, sampled at 5 $\mathrm{\mu s}$ intervals, in two polarisations. Real-time processing generally ensures that the 400 MB of data per second are reduced to manageable rates both for storage and further processing.

The ARTEMIS servers can perform in real-time all the operations necessary to discover short duration radio pulses from pulsars and fast transients, thanks to a modular software structure operating in a C++ scalable framework developed at the University of Oxford. ARTEMIS includes processing modules for receiving the data, further channelisation in finer frequency channels using a polyphase filter, generation of Stokes parameters, excision of terrestrial radio frequency interference, temporal integration, real-time brute force de-dispersion using typically at least 2000 trial dispersion measure (DM) values and detection of interesting signals, in high-throughput CPU and GPU code. The GPU processing allows a full search in DM space up to a maximum DM as defined by arguments related to interstellar scattering and with the necessary DM resolution (\cite[Armour et~al. 2012]{akg+12}). Simulations show that typically a few thousand trial DMs are sufficient to optimally sample the DM range within all LOFAR bands.

At present, GPU based ARTEMIS hardware exists at the international
stations in the UK and France (Chilbolton and Nan\c{c}ay). This allows
these stations to take advantage of the most powerful technique of
rejecting signals of terrestrial origin: the direct comparison of detections between distant sites, also referred to as anti-coincidence testing. Both these stations can process the full available 48 MHz of beamformed data. In addition two test installations exist at the stations in J\"{u}lich (Germany) and Onsala (Sweden), each able to process 12 MHz of sky bandwidth.

\section{GPU advances in ARTEMIS}

The first GPU project with ARTEMIS has been to focus on the de-dispersion step in the transients pipeline. This is clearly the most computationally intensive part of the pipeline and a preliminary study of different technologies has proved GPUs to be the most suitable accelerators for de-dispersion.

\begin{figure}[t]
\begin{center}
\includegraphics[width=0.9\textwidth]{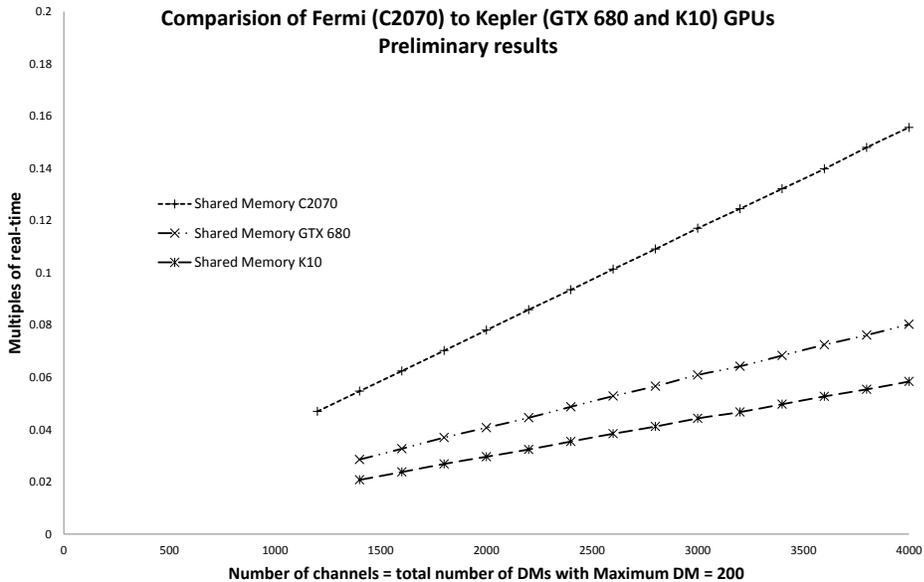}
\caption{Plot of the number of channels versus execution time as multiples of real-time. Here we hold the total number of trial dispersion searches equal to the number of channels.}
\label{fig2}
\end{center}
\end{figure}

Figure\,\ref{fig2} illustrates the speed-up that has been achieved using GPUs and compares the fastest GPU algorithm on NVIDIA Fermi and Kepler based hardware. This work has led to development of the fastest de-dispersion code in the world, with rough estimation that the new Kepler algorithm is over 2.5 times faster than previous Fermi code. The Shared Memory algorithm in Figure\,\ref{fig2} has a maximum DM limit of 100. With this limit a throughput of approximately 12~Gb/s of LOFAR data can be achieved by the de-dispersion algorithm on a Kepler K10 GPU. This DM limit can be relaxed by using a different algorithm, called the L1 algorithm, which can still comfortably process a 3.2~Gb/s single station LOFAR data stream up to a DM of 500 on a K10. It is estimated that with 10 NVIDIA K10 GPUs the L1 algorithm could process a 127-beam Tied-Array with 2000 channels in each pencil beam to a maximum DM of 250.

\end{document}